\documentstyle[12pt]{article}
\newcommand{\Pramana}{Pram\={a}\d{n}a\,}
\newcommand{\Alfven}{Alfv\'{e}n \,}
\newcommand{\Painleve}{Painlev\'{e} \,}
\newcommand{\Henon}{H\'{e}non}
\newcommand{\Pecseli}{P\'{e}cseli\,}

\newcommand{\Mjolhus}{Mj\o\,$\!$lhus\,}

\newcommand{\Schrodinger}{Schr\"{o}dinger}
\newcommand{\Ronnmark}{R\"{o}nmark\,}
\newcommand{\Naslund}{N\"{a}slund\,}

\newcommand{\eqspace}{\\[-1.11mm]}
\newcommand{\eqtwospace}{\\[-6.78mm]}
\newcommand{\xhat}{\hat{x}}

\newcommand{\zhat}{\hat{z}}

\newcommand{\PiE}{{\displaystyle \Pi}_{\scriptscriptstyle E}}
\newcommand{\PiN}{\Pi_{\scriptscriptstyle N}}

\newcommand{\omegaHo}{\omega_{{\scriptscriptstyle H}0}}
\newcommand{\omegao}{\omega_{0}}

\newcommand{\Omegaeo}{\Omega_{e0}}

\newcommand{\omegapeo}{\omega_{pe0}}

\newcommand{\VA}{V_{\scriptscriptstyle A}}
\newcommand{\Vg}{V_{g}}
\newcommand{\Bo}{B_{0}}

\newcommand{\vecBo}{\vec{B}_{0}}

\newcommand{\ko}{k_{0}}

\newcommand{\Do}{D_{0}}

\newcommand{\vte}{v_{te}}

\newcommand{\VM}{V_{\scriptscriptstyle M}}

\newcommand{\Cs}{C_{s}}

\newcommand{\Te}{T_{e}}

\newcommand{\mi}{m_{i}}
\newcommand{\me}{m_{e}}

\begin{document}
\baselineskip=20pt
\title
{H\'{E}NON--HEILES HAMILTONIAN FOR COUPLED \\[0.8ex]
UPPER--HYBRID AND MAGNETOACOUSTIC \\[0.8ex]
WAVES IN MAGNETIZED PLASMAS}
\author
{\\\\\\N. N. Rao\\
Theoretical Physics Division\\
Physical Research Laboratory\\
Navrangpura, Ahmedabad 380009\\
INDIA\\\\\\}
\date{6 March 1995}
\maketitle

\vspace{4ex}

\begin{abstract}
\baselineskip=20pt

\par We show that the coupled mode equations for the stationary
propagation of upper--hybrid and magnetoacoustic
waves in magnetized electron--ion plasmas with negative group dispersion
can be exactly derived from the
generalized \Henon--Heiles Hamiltonian. The parameter regimes for the
integrable cases of the coupled mode equations
have been explicitly obtained. For
positive group dispersion of the upper--hybrid waves,
the relevant governing equations lead to a novel Hamiltonian
where the kinetic energy is not positive definite.
\end{abstract}

\vskip .1in
\baselineskip=20pt
\newpage

\setlength{\parskip}{2ex}
\setlength{\parindent}{2em}

\noindent {\Large \bf 1. Introduction}
\vspace{2ex}

\par Amplitude modulated upper hybrid waves in a magnetized
electron--ion plasma are known to be governed by a \Schrodinger--like
equation wherein the potential is given in terms of the associated
low--frequency number density perturbations [1--3].
For small amplitudes, the latter are
governed by a linear wave equation driven by the ponderomotive
force of the high--frequency upper--hybrid waves. On the other
hand, for finite amplitudes, the density perturbations are governed by
a driven nonlinear Boussinesq equation [4] which is coupled to the
\Schrodinger \, equation. For uni--directional propagation,
the (driven) Boussinesq equation reduces to the
well--known (driven) Korteweg--de Vries (K--dV) equation. For stationary
propagation of the coupled waves, the time--dependent
\Schrodinger--Boussinesq (or,
K--dV) equations give rise to a coupled system of nonlinear ordinary
differential equations which can be derived from a Hamiltonian. While
special analytical solutions valid in some specific regions of the
allowed parameter space have been obtained [5], the question of
the complete integrability of the associated Hamiltonian for arbitrary
boundary conditions has not yet been answered.

\par We show, in this Letter, that the coupled mode equations
for the upper--hybrid and magnetoacoustic waves with negative
group dispersion can be exactly reduced to the generalized
\Henon--Heiles equations which are extensively studied in the
field of Nonlinear Dynamics [6--8]. We thereby
obtain explicitly the parameter regimes for the integrability of the
associated Hamiltonian. For the case of positive group dispersion, the
coupled mode equations are derivable from a novel kind of Hamiltonian
having indefinite kinetic energy.

\vspace{4ex}
\noindent {\Large \bf 2. Governing Equations}
\vspace{2ex}

\par  We consider the one--dimensional propagation along
$\xhat$--direction of the high--frequency upper--hybrid waves in a
homogeneous,
magnetized electron--ion plasma with the external magnetic field
($\vecBo$) along the $\zhat$--direction, that is, $\vecBo=\Bo \zhat$.
For normal modes, the upper--hybrid wave frequency ($\omegao$)
and the wavenumber ($\ko$) are related by the linear
dispersion relation [9],
\eqspace
$$\omegao^2 = \omegaHo^2 + \frac{3 \omegapeo^2 \ko^2 \vte^2}
{\omegao^2 -4 \Omegaeo^2} \,, \eqno(1)$$
\eqspace
where $\omegapeo = (4 \pi n_0 e^2 /\me)^{1/2}$ is the electron plasma
frequency, $\Omegaeo= e\Bo/\me c$ is the electron gyro--frequency,
$\vte=(\Te/\me)^{1/2}$ is the electron thermal speed, $\omegaHo =
(\omegapeo^2 + \Omegaeo^2)^{1/2}$ is the upper--hybrid frequency, and
all the other symbols have their usual meanings [4]. In the long
wavelength limit, Eq. (1) can be approximated by
\eqspace
$$\omegao =\omegaHo + \frac{1}{2} \, \Do \ko^2\,, \eqno(2) $$
\eqtwospace
where
\eqtwospace
$$\Do \equiv \frac{\partial^2 \omegao}{\partial \ko^2} = \frac{3
\omegapeo^2 \vte^2}{\omegaHo (\omegapeo^2 - 3 \Omegaeo^2)} \,,
\eqno(3) $$
\eqspace
denotes the group dispersion coefficient for the upper--hybrid waves.
Clearly, the latter have positive (negative) dispersion for plasma
parameters such that $\omegapeo^2 > 3 \Omegaeo^2$\, ($\omegapeo^2 < 3
\Omegaeo^2$).

\par For nonlinear propagations, the slowly varying complex amplitude
$E(x,t)$ of the upper--hybrid wave electric field is governed by a
\Schrodinger \, equation of the form [2,4],
\eqspace
$$i\, \left( \frac{\partial E}{\partial t} + \Vg \, \frac{\partial E}
{\partial x} \right) \, + \, \frac{\Do}{2}\, \frac{\partial^2 E}{
\partial x^2} \,=\, \mu \,\omegaHo N E\,, \eqno(4)$$
\eqspace
where, $N=\delta n_e/n_0$ is the normalized low--frequency density
perturbation, $\Vg\equiv \partial \omegao/\partial \ko = \ko \Do$
denotes the group velocity and $\mu= \frac{1}{2} (\omegapeo^2 +2
\Omegaeo^2)/(\omegapeo^2 + \Omegaeo^2)$.

\par For finite amplitudes, the low--frequency density perturbation
($N$) is governed by a Boussinesq or a K--dV type of nonlinear
equation which is driven by the ponderomotive force of the
high--frequency upper--hybrid waves. The driven Boussinesq or the
K--dV equation can be derived from the low--frequency fluid equations
(for the electrons and the ions) which are coupled to the Maxwell
equations. Omitting the details of the derivation
[4], we write below the driven Boussinesq equation in the
form,
\eqspace
$$\frac{\partial^2 N}{\partial t^2} - \VM^2\, \frac{\partial^2 N}
{\partial x^2} -\theta^2 \, \frac{\partial^4 N}{\partial x^4} -a^2
\frac{\partial^2}{\partial x^2} \left(N^2\right) = \eta^2
\frac{\partial^2}{\partial x^2} \left( \frac{|E|^2}{16 \pi n_0 \Te}
\right) \,, \eqno(5)$$
\eqspace
where, $\VM=(\VA^2 +\Cs^2)^{1/2}$ is the magnetoacoustic speed,
$\VA=(\Bo^2/4 \pi n_0 \mi)^{1/2}$ is the \Alfven speed,
$\Cs=(\Te/\mi)^{1/2}$ is the ion--acoustic speed, $\theta = c
\VA/\omegapeo$, $a^2=(3 \VA^2 + 2 \Cs^2)/2$, and $\eta= \omegaHo
\Cs/\omegapeo$. Note that in the linear limit for normal modes, Eq.
(5) gives,
\eqspace
$$\omega^2 = \VM^2 k^2 -\theta^2 k^4\,, \eqno(6)$$
\eqspace
which is the linear dispersion relation for the magnetoacoustic modes
in the long wavelength regime. Equation (5) describes the
bi--directional propagation of the nonlinear magnetoacoustic waves
driven by the upper--hybrid ponderomotive force.

\par On the other hand, for uni--directional propagation, Eq. (5) can
be reduced to the form,
\eqspace
$$\frac{\partial N}{\partial t} + \VM\, \frac{\partial N}
{\partial x} +\frac{\theta^2}{2 \VM}
 \, \frac{\partial^3 N}{\partial x^3} +\frac{a^2}{\VM}
N \frac{\partial N}{\partial x}  = -\frac{\eta^2}{2 \VM}
\frac{\partial}{\partial x} \left( \frac{|E|^2}{16 \pi n_0 \Te}
\right) \,, \eqno(7)$$
\eqspace
which is the driven K--dV equation. For normal modes, Eq. (7) yields
the dispersion relation, $\omega= \VM k - \theta^2 k^3/2\VM$ which
follows also from Eq. (6) in the small wavenumber limit.

\par For stationary propagation, the wave fields are expressed in form,
\eqspace
$$E(x,t)= E(\xi)\, \exp \left[ i\left\{ X(x) + T(t) \right\}
\right]\,, \eqno(8.1)$$
\eqtwospace
$$N(x,t)=N(\xi)\,, \eqno(8.2)$$
\eqspace
where $\xi=x-Mt$ represents  the coordinate in the
stationary frame whose speed is
determined by the free parameter $M$; the functions $X(x)$ and $T(t)$
are introduced in Eq. (8.1) to account for the possible shifts in the
wavenumber as well as in the wave frequency due to the nonlinear
interactions. Using Eqs. (8.1) and (8.2) in Eq. (4), we obtain,
\eqspace
$$\Do\, \frac{d^2 E}{d \xi^2} = \lambda E + b^2 NE\,, \eqno(9)$$
\eqspace
where $\lambda= 2 \delta+ (M^2-\Vg^2)/\Do$ is
the nonlinear shift parameter,  $\delta=dT/dt$
denotes the shift in the wave frequency and $b^2=2\,\mu\,\omegaHo$.

The stationary
governing equation for the
density perturbations ($N$) is obtained from Eq. (5) or (7) as,
\eqspace
$$\theta^2 \, \frac{d^2 N}{d \xi^2} = f N -a^2 N^2
-\eta^2 \, \frac{E^2}{16 \pi n_0 \Te}\,, \eqno(10)$$
\eqspace
where,
\eqspace
$$ f= \left\{
\begin{array}{llr}
M^2-\VM^2\,,  & \mbox{for the driven Boussinesq Eq. (5),} \\[3.0ex]
2\VM(M- \VM)\,, & \mbox{for the driven K--dV Eq. (7).}
\end{array} \right. \eqno(11)$$

\par Equations (9) and (10) are the relevant coupled mode equations for
the stationary propagation of upper--hybrid and
magnetoacoustic waves. We have presented elsewhere [5,10] different
classes of exact analytical solutions which either
use special boundary conditions or are valid in limited regions of the
allowed parameter space. In the next section, we reduce these
equations to the generalized \Henon--Heiles form and thereby determine
the integrable parameter regimes.

\vspace{4ex}
\noindent {\Large \bf 3. Reduction to \Henon--Heiles equations}
\vspace{2ex}

\par In order to reduce the coupled mode equations (9) and (10) to the
generalized \Henon--Heiles form, we note that the parameters
$\lambda$, $\Do$ and $f$ can have either signs.
We normalize $N$ with respect to $-2 \Do D / b^2 $ and
$E^2$ with respect to $16 \pi n_0 \Te (2 |\Do| D/b^2)$, and obtain,
\eqspace
$$\frac{d^2 E}{d \xi^2} = \frac{\lambda}{\Do} E \,
  - \frac{2 \eta^2}{\theta^2}\, E N\,, \eqno(12)$$
\eqtwospace
$$\frac{d^2 N}{d \xi^2} = \frac{f}{\theta^2}\, N
 + p\, \frac{2 a^2 \eta^2\,|\Do|}{b^2 \theta^4} N^2
 + p\, \frac{\eta^2}{\theta^2}\, E^2\,, \eqno(13)$$
\eqspace
where $p=-1$ for negative dispersion ($\Do<0$) and $p=+1$ for positive
dispersion ($\Do>0$) of the upper--hybrid waves. We consider below
these two cases separately.

\vspace{2ex}
\noindent {\bf (A) \,\,Negative dispersion \, ($\Do <0$)}
\vspace{2ex}

\par For upper--hybrid waves with negative group dispersion
($\omegapeo^2 < 3 \Omegaeo^2$), the coupled mode equations (12) and
(13) become,
\eqspace
$$\frac{d^2 E}{d \xi^2} =- A E - 2 D E N\,, \eqno(14)$$
\eqtwospace
$$\frac{d^2 N}{d \xi^2} = -B N - C N^2 - D E^2\,. \eqno(15)$$
\eqtwospace
where,
\eqtwospace
$$A= \frac{\lambda}{|\Do|}\,,\,\,\,\,\,
B= - \frac{f}{\theta^2}\,,\,\,\,\,\,
C= \frac{2 a^2 \eta^2 |\Do|}{b^2 \theta^4}\,,\,\,\,\,\,
D= \frac{\eta^2}{\theta^2}\,. \eqno(16)$$
\eqspace
Equations (14) and (15) can be derived from the Hamiltonian,
\eqspace
$$H_+ = \frac{1}{2} \left( \PiE^2 + \PiN^2 \right) +\frac{1}{2} \left(
AE^2 +B N^2\right) + \left( \frac{1}{3} CN^3 +DNE^2\right) \,,
\eqno(17)$$
\eqspace
where $\PiE \equiv dE/d\xi$ and $\PiN \equiv dN/d\xi$ are,
respectively, the \lq\lq canonical momenta\rq\rq\,
conjugate to $E$ and $N$.

\par Clearly, by treating $E$ and $N$ as the
spatial coordinates and $\xi$ as the temporal coordinate, Eq. (17) may
be considered as the Hamiltonian for the two--dimensional motion of a
pseudo--particle of unit mass. In fact, $H_+$ is identically the same as
the generalized \Henon--Heiles Hamiltonian [7,11] which has been
extensively studied in the field of Nonlinear Dynamics.
Since the stationary coordinate ($\xi$) does not explicitly appear in
Eq. (17), the associated potential is conservative and hence the
Hamiltonian $H_+$ is an integral of motion. For the two--dimensional
motion, the system is completely integrable provided there exists the
second integral of motion which is in involution with the Hamiltonian
[12]. On the other hand, in recent years, the so--called \lq\lq
\Painleve Analysis\rq\rq\, has been extensively used to obtain the
parameter regimes wherein low--dimensional Hamiltonian systems may
be completely integrable [7,8].
In particular, it is well--known that the Hamiltonian (17) is
completely integrable for the following three sets of parameter
values :
\eqspace
$$\hspace{-41.5ex}
\mbox{(a)}\,\,\, A=B \,, \,\,\,\,\,\,\,\, C=D \eqno(18.1)$$
$$\hspace{-36.5ex}
\mbox{(b)}\,\,\, 16\, A=B\,, \,\,\,\,\,\,\,\, C=16\, D \eqno(18.2)$$
$$\hspace{-26.0ex}
\mbox{(c)}\,\,\, \mbox{arbitrary}\,\,\, A\,\,\, \mbox{and}\,\,\,
B, \,\,\,\,\,\,\,\, C=6\, D\,.\eqno(18.3)$$
\eqspace
The associated second integrals of
motion for the above parameters have been summarized in Ref. [8].
Furthermore, there are indications from general considerations that
these are  possibly the only integrable cases of the generalized
\Henon--Heiles Hamiltonian [11]. Thus, a necessary condition for the
integrability of Eqs. (14) and (15) seems to be
that the nonlinear terms should have the same signs.

\par To obtain explicitly the various
plasma parameters for integrability, we
shall first consider the case when Eq. (10) together with (11)
corresponds to the driven
Boussinesq equation (5). In terms of the dimensionless parameters
defined by $\alpha = \omegapeo/\Omegaeo$,\, $\beta= (\Cs/\VA)^2$\,
and $\gamma= \vte/c$, the above relations can be written in the form,
\eqspace
$$3 \alpha^4 \gamma^2 (3 + 2 \beta)= \nu ( 2+ \alpha^2)(3 -\alpha^2)
\,, \eqno(19)$$
\eqtwospace
and
\eqtwospace
$$3 \alpha^4 \gamma^2 (1 + \beta)(1-M^2)= \nu \Lambda
(1+ \alpha^2)(3 -\alpha^2) \,, \eqno(20)$$
\eqspace
where $\Lambda \equiv \lambda/\omegaHo$ is the normalized nonlinear
shift parameter, $M$ is the Mach number normalized
with respect to $\VM$ and $\nu$ takes values 1, 6, or 16. Note that
the parameter $\beta$ is essentially the usual plasma beta, that is,
the ratio of the thermal pressure to the magnetic field pressure. For
integrability of the coupled
Eqs. (14) and (15), both the conditions (19) and
(20) should be simultaneously satisfied for the case
when $\nu =1$ or 16, whereas only the condition (19) needs to be
satisfied when $\nu=6$.

\par Defining $\Gamma^2 \equiv \gamma^2/\nu$, Eq. (19) can be written
in the form,
\eqspace
$$\Gamma^2 = \frac{( 2+ \alpha^2)(3 -\alpha^2)}{3 \alpha^4
(3+2\beta)} \,. \eqno(21)$$
\eqtwospace
Note that $\Gamma^2$ remains positive definite since $\alpha^2<3$ is
satisfied for negative group dispersion. Substituting for $\gamma^2$
from Eq. (19) into Eq. (20), we get,
\eqspace
$$\Delta \equiv \frac{1-M^2}{\Lambda} = \frac{(1+\alpha^2)(3 +
2\beta)}{(2 + \alpha^2)(1 + \beta)}\,. \eqno(22)$$
\eqspace
Figure (1) shows a plot of $\Gamma^2$ as a function of $\alpha^2$ from
Eq. (21) for different values of $\beta$.
The value $\beta=0$ corresponds to the cold plasma case whereas large
values of $\beta$ correspond to weakly magnetized plasmas.
For $\nu=6$, the system of equations (14)
and (15) is completely integrable for parameters given by Figure 1
and which satisfy $\Gamma^2=\beta/\alpha^2 \nu$ since $\beta\equiv
\alpha^2 \gamma^2$. Note that for $\nu=6$, any arbitrary values of
$M$ and $\Lambda$ are admissible. On the other hand, Figure
(2) gives a plot of $\Delta$ as a function of $\alpha^2$ from Eq. (22)
for different values of $\beta$. For $\nu=1$ and 16,  the corresponding
values of the parameter
$\gamma$ for integrability are given by Figure 1 whereas the
parameters $M$ and $\Lambda$ are no longer arbitrary but are
related by Eq. (22). Since the right--hand side of Eq. (22) is
positive definite always, it follows that for
positive (negative) values of the frequency shift parameter
$\Lambda$, the governing equations (14) and (15) are integrable for
sub--magnetoacoustic, that is, for $M<1$ (super--magnetoacoustic,
$M>1$) values of the Mach number $M$.

\par The parameter values for integrability for the
driven K--dV case in Eqs. (10) and (11) can similarly be obtained. In
fact, it follows by inspection that Eqs. (21) and (22) hold good in
this case also provided the factor $(1-M^2)$ in the latter
is replaced by $2(1-M)$.
The Mach number ($M$) regimes for integrability are, therefore,
qualitatively the same as in the driven Boussinesq case.

\vspace{2ex}
\noindent {\bf (B)\,\, Positive dispersion \, ($\Do >0$)}
\vspace{2ex}

\par For $\omegapeo^2 > 3 \Omegaeo^2$, that is, for $\alpha^2>3$, the
upper--hybrid waves have positive group dispersion and the coupled
mode equations (12) and (13) become,
\eqspace
$$\frac{d^2 E}{d \xi^2} = A E - 2 D E N\,, \eqno(23)$$
\eqtwospace
$$\frac{d^2 N}{d \xi^2} = -B N + C N^2 + D E^2\,, \eqno(24)$$
\eqspace
where the coefficients $A$, $B$, $C$ and $D$ are defined, as earlier,
by Eqs. (16). The coupled equations (23) and (24)
are derivable from the Hamiltonian,
\eqspace
$$H_- = \frac{1}{2} \left( \PiE^2 - \PiN^2 \right) - \frac{1}{2} \left(
AE^2 + B N^2\right) + \left( \frac{1}{3} CN^3 +DNE^2\right) \,,
\eqno(25)$$
\eqspace
where the canonical momenta, for the present case, are given by $\PiE
\equiv dE/d\xi$ and $\PiN \equiv - dN/d\xi$. As earlier, the
Hamiltonian $H_-$ is an integral of motion corresponding to Eqs. (23)
and (24).

\par The Hamiltonians
$H_+$ and $H_-$ given, respectively, by Eqs. (17) and (25) differ from
each other by sign changes for the terms containing $\PiN^2$,
$E^2$ and $N^2$. The sign changes for the terms containing
$E^2$ and $N^2$ are trivial and can, in fact, be absorbed by
redefining the coefficients $A$ and $B$. However, the sign change
for the quadratic term in $\PiN$ is indeed
significant. For, unlike the case of the
Hamiltonian $H_+$, the \lq\lq kinetic energy\rq\rq \, term in the
Hamiltonian $H_-$ is not positive definite.
The latter is in contrast to the usual Hamiltonian systems of
classical dynamics where the kinetic energy is positive definite
always. An important consequence of this difference  is that systems
with indefinite kinetic energy need not necessarily have bounded
motions around those points where the potential energy has a minimum.
In fact, for such systems, both the canonical momenta can
simultaneously increase or decrease but still keeping the Hamiltonian
an integral of motion. Furthermore, the usual stability
theorems developed in classical dynamics may not be directly
applicable to such cases. It is therefore expected
that the qualitative nature of the solutions in both the cases should
be different. Hamiltonians with indefinite kinetic energy [13]
and having different kinds of potential functions are known to
arise in many problems dealing with the nonlinear evolution of
the modulational instability of an high--frequency wave coupled to a
suitable low--frequency wave. We have reported earlier [5,10] some
special classes of exact analytical solutions of the coupled equations
(23) and (24). However, the question of the complete integrability of
such Hamiltonians, in general, and that
of $H_-$ given by Eq. (25), in particular, by either the standard
\Painleve analysis or otherwise is still open.

\newpage
\vspace{4ex}
\noindent {\Large \bf 4. Conclusions}
\vspace{2ex}

\par To conclude, we have shown that the stationary governing
equations for modulated upper--hybrid waves coupled to magnetoacoustic
waves in a magnetized plasma with negative group dispersion are
derivable from the generalized \Henon--Heiles Hamiltonian of nonlinear
dynamics. The parameter regimes for the integrability of the
associated Hamiltonian have been explicitly obtained.
For the case of
upper--hybrid waves with negative group dispersion, the equations
give rise to a novel type of Hamiltonian with indefinite kinetic
energy. The \Painleve analysis as well as the integrability of the
latter remains to be investigated.

\par The results of the present investigation should have a bearing
on the possible parameter regimes for the occurrence of upper--hybrid
turbulence in magnetized plasmas [3,14,15]. For example, magnetized
plasma turbulence has been invoked as a
possible source for the narrow--band, non--thermal continuum radiation
observed in the Earth's magnetosphere [16]. In the present work, we
have explicitly obtained those parameter regimes where the stationary
wave fields are non--chaotic and hence can give rise to soliton--like
coherent nonlinear structures [17].

\vspace{4ex}
\noindent {\Large \bf Acknowledgement}

\par The author thanks Dr. B.R. Sitaram for useful discussions.
\vspace{2ex}

\newpage

\vspace{4ex}
\noindent {\Large \bf References}
\vspace{2ex}

\noindent\,\,[1]\,\, A.N. Kaufman and L. Stenflo,
Physica Scripta {\bf 11} (1975) 269.

\noindent\,\,[2]\,\, M. Porkolab and M.V. Goldman,
Phys. Fluids {\bf 19} (1976) 872.

\noindent\,\,[3]\,\, D. ter Haar, Physica Scripta {\bf T2/2} (1982) 522.

\noindent\,\,[4]\,\, N.N. Rao, J. Plasma Phys. {\bf 39} (1988) 385.

\noindent\,\,[5]\,\, N.N. Rao, J. Phys. {\bf A22} (1989) 4813.

\noindent\,\,[6]\,\, M. \Henon \, and C. Heiles,
Astron. J. {\bf 69} (1964) 73.

\noindent\,\,[7]\,\, Y.F. Chang, M. Tabor and J. Weiss,
J. Math. Phys.  {\bf 23} (1982) 531.

\noindent\,\,[8]\,\, M. Lakshmanan and R. Sahadevan,
Phys. Rep. {\bf 224} (1993) 1.

\noindent\,\,[9]\,\, K.B. Dysthe, E. \Mjolhus, H.L. \Pecseli and L.
Stenflo, Plasma Phys. Contro. Fusion\\
\indent \indent {\bf 26} (1984) 443.

\noindent [10]\,\, N.N. Rao, J. Phys. {\bf A24} (1991) L993.

\noindent [11]\,\, A.P. Fordy, Physica {\bf D52} (1991) 204.

\noindent [12]\,\, A.J. Lichtenberg and M.A. Lieberman,
{\em Regular and Stochastic Motion}\\
\indent \indent  (Berlin, Springer, 1983) p. 23.

\noindent [13]\,\, N.N. Rao, B. Buti and S.B. Khadkikar,
\Pramana -- J.  Phys. {\bf 27} (1986) 497.

\noindent [14]\,\, Z.Z. Kasymov, E. \Naslund, A.N. Starodub and L.
Stenflo, Physica Scripta {\bf 31} (1985) 201.

\noindent [15]\,\, P.K. Shukla, R. Fedele and U. de Angelis,
Phys. Rev. {\bf A31} (1985) 517.

\noindent [16]\,\, P.J. Christiansen, J. Etcheto, K. \Ronnmark and L.
Stenflo, Geophys. Res. Letts.\\
\indent \indent {\bf 11} (1984) 139.

\noindent [17]\,\, I. Mori and K. Ohya,
Phys. Rev. Letts. {\bf 59} (1987) 1825.

\newpage
\vspace{10ex}
\begin{center}
{\Large \bf Figure Captions}
\end{center}

\vspace{10ex}
\begin{description}
\item{\noindent \bf Figure 1. \,\,\,\,\,} Parameters $\Gamma$
and $\alpha$ for
the integrability of the coupled mode equations (14) and (15)
with negative group
dispersion. $\beta=0$ corresponds to the case of cold plasmas whereas
large $\beta$ corresponds to the weakly magnetized plasmas. For
$\nu=6$, parameters $M$ and $\Lambda$ can  independently
take arbitrary values for integrability.\\\\

\item{\noindent \bf Figure 2. \,\,\,\,\,} Parameters $\Delta$ and
$\alpha$ for the integrability of the coupled mode equations (14) and
(15) with negative group dispersion.
$\beta=0$ corresponds to the case of cold plasmas whereas
large $\beta$ corresponds to the weakly magnetized plasmas. For
$\nu=1$ or 16, the corresponding values of $\gamma=\sqrt{\nu}\, \Gamma$
are to be obtained from Figure 1. Note that unlike the $\nu=6$ case,
both the parameters $M$ and $\Lambda$ are not arbitrary for $\nu=1$ or
16, but are related by Eq. (22).

\end{description}

\end{document}